# REALIZATION OF A DIAMOND BASED HIGH DENSITY MULTI ELECTRODE ARRAY BY MEANS OF DEEP ION BEAM LITHOGRAPHY

F. Picollo[1,2,3], A. Battiato[2,1,3], E. Bernardi[2,1,3], L. Boarino[4], E. Enrico[4], J. Forneris[2,1,3], D. Gatto Monticone[2,1,3], P. Olivero[2,1,3]

[1] Istituto Nazionale di Fisica Nucleare (INFN), Sezione di Torino, Torino, Italy

[2] Physics Department and "NIS" Inter-departmental Centre, University of Torino, Torino, Italy

[3] Consorzio Nazionale Inter-universitario per le Scienze fisiche della Materia (CNISM), Sezione di Torino, Torino, Italy

[4] Nanofacility Piemonte, National Institute of Metrologic Research (INRiM), Torino, Italy

**Keywords**: Deep Ion Beam Lithography, diamond, biosensors, multi electrode array

**ABSTRACT**

In the present work we report about a parallel-processing ion beam fabrication technique whereby high-density sub-superficial graphitic microstructures can be created in diamond. Ion beam implantation is an effective tool for the structural modification of diamond: in particular ion-damaged diamond can be converted into graphite, therefore obtaining an electrically conductive phase embedded in an optically transparent and highly insulating matrix.

The proposed fabrication process consists in the combination of Deep Ion Beam Lithography (DIBL) and Focused Ion Beam (FIB) milling. FIB micromachining is employed to define micro-apertures in the contact masks consisting of thin (<10 μm) deposited metal layers through which ions are implanted in the sample. A prototypical single-cell biosensor was realized with the above described technique. The biosensor has 16 independent electrodes converging inside a circular area of 20 μm diameter (typical neuroendocrine cells size) for the simultaneous recording of amperometric signals.

**1. Introduction**

In the last decade diamond has attracted interest for the development of electronic devices with promising performances [1] owing to its extreme electrical properties. Significant effort has been made to optimize the interfacing of diamond with conventional electronics, resulting in the development of techniques for the fabrication of electrical contacts and electrodes in this material. Different approaches have been adopted, ranging from surface processing such as metallization [2] or hydrogen termination [3, 4], to bulk doping achieved by ion implantation [5]. Moreover, high power pulsed laser was employed to promote the diamond graphitization both on the surface and in the bulk [6].

Besides the above-mentioned techniques, ion-beam-induced graphitization of diamond has been extensively investigated with Deep Ion Beam Lithography (DIBL) [7, 8]. This approach takes advantage of the metastable nature of diamond, which can be converted into the stable allotropic form of carbon in ambient temperature and pressure conditions (i.e. graphite) by inducing high defect concentration in the lattice and by subsequently processing the material via thermal annealing [9]. The damaging with energetic ions in matter occurs mainly at the end of ion range, where the cross section for nuclear collisions is strongly enhanced [10], while the effects of electronic energy loss can be neglected in this material. In order to connect the buried implanted structures to the sample surface, a three-dimensional masking technique was developed to modulate the penetration depth of the ions from their range in the unmasked material up to the sample surface with increasing thickness of stopping material [7, 8]. The permanent conversion of ion-implanted diamond to a graphite-like phase upon thermal annealing at high temperature (>900 °C) occurs when a critical damage density (usually referred to as "graphitization threshold") is overcome. Such threshold value has been estimated as $9\times10^{22}$ vacancies cm$^{-3}$ [11]. MeV ion beams focused to micrometric spot sizes have been employed in DIBL and opened the way to the fabrication of micro-structures in diamond.

Ion beam lithography in diamond was extensively applied for the fabrication of a broad range of devices: waveguides [12 - 14], photonic structures [15 - 17], micromechanical resonators [19 - 20]. The possibility of creating graphitic conductive regions allowed the fabrication of infrared radiation emitters [21], field emitters [22] and ionizing radiation detectors [23 - 24]. Moreover, it is worth mentioning that diamond offers an intrinsic biocompatibility, a property that is functional for the realization of cellular biosensors [25].

All of the above-mentioned techniques are versatile tools for diamond modification but offer a spatial resolution limited to few micrometers due to the high currents necessary to implant the samples at the desired fluences. In the present paper we report on a parallel three-dimensional lithographic technique based on the combination of broad-beam DIBL with contact masking by means of metallic layers microfabricated by Focused Ion Beam machining.

## 2. Experimental

The sample consists of a commercial synthetic single-crystal diamond grown by chemical vapor deposition (CVD) by ElementSix. The diamond is $3\times3\times0.5$ mm$^3$ in size and it is classified as type IIa ("optical grade") with substitutional nitrogen and boron concentrations lower than 1 ppm and 0.05 ppm, respectively. The sample is cut along the 100 crystal direction and it is optically polished on the two opposite large faces.

Our DIBL technique in diamond is based on the implantation with MeV ion beams through metal masks suitably microfabricated by Focused Ion Beam (FIB).

The sample micro-machining consisted of the following fabrication steps.

Firstly, a uniform 4-µm-thick copper film was deposited directly onto the diamond surface by thermal evaporation.

Subsequently, a high-resolution mask, which defines the geometry of the implanted graphitic structures, was realized in the above-mentioned layer. The mask apertures were performed by means of FIB milling. The instrument employed is a dual beam FEI Quanta 3D™ equipped with Nanometer Pattern Generation System from J.C. Nabity. It is essential to obtain buried channels with surface-exposed end-points which will act as multiple bio-sensing electrodes for cellular *in vitro* recordings, and at the sample periphery, which will provide contacts for chip-bonding. Therefore, the milling process must produce variable thickness holes which act as graded implantation masks [7, 8]. Such configuration is achievable by opportunely tuning the milling dose during the FIB machining: the employed instrument is equipped with NPGS software, a dedicated environment for the delineation of complex structures [26]. As shown in Fig.1 highly resolved 3D metal masks for ion beam lithography were fabricated. Typical dimensions of the milled holes were width of ~2.2 μm and length of ~200-235 μm. A protective thin layer (~200 nm) was leaved on the bottom of the milled aperture in order to avoid the superficial damage induced by Ga beam. This strategy was defined because, despite having a good control of the milling dose during the FIB machining process, it was not possible to obtain a complete removal of the metal without partially milling the diamond surface.

Graphitic channels buried into the diamond substrate were fabricated by ion implantation through the previously-described masks using a 1.8 MeV $He^+$ broad-beam (~5 $mm^2$ spot size) with an ion current of ~500 nA. The ion fluence was ~$2\times10^{17}$ $cm^{-2}$ and the sample was implanted at room temperature. Ion implantation was performed by using the 60° beam line at the AN2000 facility of the INFN National Laboratories of Legnaro (I).

According to SRIM2013 Monte Carlo simulation [27], these irradiation conditions are suitable for producing a vacancy density profile with a damage peak well above the graphitization threshold (i.e. $9\times10^{22}$ $cm^{-3}$) [11].

SRIM simulations were carried out selecting the "Detailed calculation with full damage cascades" mode, by setting a displacement energy value of 50 eV for diamond [28]. It is worth noting that the vacancy density profile was obtained by modeling the effect of cumulative ion implantation on the number of vacancies with a simple linear approximation, which does not take into account for more complex processes such as self-annealing, ballistic annealing and defect interaction, and assuming a linear dependence of the conductivity on the vacancy density. Therefore, the estimated value for the graphitization threshold can only be considered as an effective parameter quantifying the induced damage density.

After ion implantation, the Cu masks were removed from the surface. The sample was then annealed in vacuum at 950 °C for 2 hours with slow heating and cooling rates (5 °C min$^{-1}$) to avoid thermal stress. The thermal treatment was performed with the purpose of converting the highly-damaged regions located at the ion end of range to a graphitic phase while removing the structural sub-threshold damage introduced in the layer overlying the above-mentioned damaged region.

The electrical characterization (two-terminal current-voltage measurement) of the channels was performed with a home-built setup consisting of a system of microprobes operating at room temperature. The microprobes were connected by suitably shielded feed-throughs to a Keithley model 614 which can supply a voltage between -3 V and +3 V and simultaneously measure the current flowing in the circuit. The maximum resolution of the instrument is 10 fA and the accuracy is typically better than 1%, depending on the used scale used and quality of the contacts. In order to improve the quality of the electrical contact between the channels and the microtips, 200 nm thick silver pads were deposited over the emerging endpoints of the graphitic micro-structures.

**3. Results**

The design of a DIBL process requires an accurate evaluation of the mask thickness and of the ion beam energy involved in the process. These parameters can be established by means of Monte Carlo numerical simulations (see Fig. 2). Using this approach we chose the above reported values (mask thickness = 4 μm; $He^+$ beam energy = 1.8 MeV) in order to fabricate conductive strips 3 μm below the surface.

As shown in Fig. 2a, the Bragg peak of the implanted $He^+$ ions is entirely located within the copper layer, thus guaranteeing a complete masking of the areas unexposed to the MeV ion irradiation. At the same time, the FIB milled regions of the metal film allowing the passage of the ions. The depth profile of the ion-induced structural damage in these areas is reported in Fig. 2b, being parameterized as the volumetric density of induced vacancies, as resulting from SRIM 2013.00 Monte Carlo code [27].

The optical micrograph presented in Fig. 3 shows the graphitic structures obtained in diamond after the thermal treatment. The realized geometry was planned in order to produce a multi-electrode-array biosensor capable of interfacing with a single cell [25]: indeed the 16 electrodes are converging inside a circular area of 20 μm diameter. It is also worth stressing that the emerging peripheral zones are laid out to easily provide contacts for subsequent chip bonding.

The spatial resolution in the fabrication of the graphitic features of the high-density multi-electrode array was achieved thanks to the micrometric resolution of the FIB mask milling. In particular, it is worth noting that the lateral straggling of MeV ions (~ 0.2 μm) does not significantly degrade the resolution of the lithographic process because, at the set fluence, laterally straggling ions do not induce a damage density high enough to determine the amorphization of the material, i.e. the thermal annealing process effectively "erases" their structural effects.

Suitably thinning Cu ramps realized on the endpoints of the FIB-milled holes in the metal mask ensured the electrical continuity of the buried channels with the surface.

The electrical properties of the 16 implanted microchannels were investigated by current-voltage measurements. An example of IV curve obtained by probing two pads located at the endpoints of a single channel is shown in Fig. 4: the IV characteristic shows an ohmic behavior and a resistance of ~10.5 MΩ is obtained. The resistivity of the conductive channel can be estimated from the geometrical dimensions of the structure. In particular the lateral dimensions (length, width) are known from optical microscopy, while the thickness was estimated from the intersection of the graphitization threshold with the SRIM-derived vacancy density profile (see Fig. 1). The obtained resistivity of the channels is ~1.9 mΩ cm, comparable with the one of common polycrystalline graphite (1.3 mΩ cm) [29].

## 4. Conclusions

In the present paper we reported on the realization of a diamond-based high-density multi-electrode array for single-cell measurements by means of a parallel ion beam lithographic technique.

The employment of DIBL with Focused Ion Beam machining on diamond offers the advantage of being parallel as far as broad MeV implantation is concerned, allowing the simultaneous definition of several graphitic structures, and it can be further improved towards sub-micrometric resolution despite the high MeV ion currents employed (~ nA) by optimizing the FIB-milling process, since it is not suffering of ion straggling.

The proposed technique allowed the realization of high-density graphitic microstructures that act as electrodes and that will be exploited in a single-cell multi-electrode biosensor, taking advantage of the chemical inertness, biocompatibility and optical transparency of diamond. The prototype was designed for the simultaneous recording of amperometric signals from single neuroendocrine cells by means of 16 independent electrodes converging inside a circular area of 20 μm diameter, with the aim of bringing new insights into the spatial mapping of secretory sites. The investigation of the micro-domain organization of neurosecretion with high spatial resolution is of paramount importance in modern neurophysiology, because the presence of highly localized areas for secretion and large areas of silent activity within the same cell is the subject of an increasing number of studies in the field [30 - 32].

**Acknowledgements**

This work is supported by the following projects: "DiNaMo" (young researcher grant, project n° 157660) by National Institute of Nuclear Physics; FIRB "Futuro in Ricerca 2010" (CUP code: D11J11000450001) funded by MIUR and "A.Di.N-Tech." (CUP code: D15E13000130003), "Linea 1A - ORTO11RRT5" projects funded by the University of Torino and "Compagnia di San Paolo". Nanofacility Piemonte is laboratory supported by the "Compagnia di San Paolo" foundation.

References

[1] R. Kalish, J. Phys. D: Appl. Phys. 40 (2007) 6467

[2] M. Werner, Semicond. Sci. Technol. 18 (2003) S41

[3] P. Ariano, A. Lo Giudice, A. Marcantoni, E. Vittone, E. Carbone, D. Lovisolo, Biosens. Bioelectron. 24 (2009) 2046

[4] T. Banno, M. Tachiki, H. Seo, H. Umezawa and H. Kawarada, Diamond Relat. Mater. 11 (2002) 387

[5] J. F. Prins, Semicond. Sci. Technol. 18 (2003) S27

[6] T.V. Kononenko, V.I. Konov, S.M. Pimenov, N.M. Rossukanyi, A.I. Rukovishnikov, V. Romano, Diamond Relat. Mater. 20 (2011) 264

[7] P. Olivero, G. Amato, F. Bellotti, O. Budnyk, E. Colombo, M. Jakšić, C. Manfredotti, Ž. Pastuović, F. Picollo, N. Skukan, M. Vannoni, E. Vittone, Diamond Relat. Mater. 18, (2009) 870–876

[8] F. Picollo, D. Gatto Monticone, P. Olivero, B. A. Fairchild, S. Rubanov, S. Prawer, E. Vittone, New J. Phys. 14 (2012) 053011

[9] S. Prawer, D. N. Jamieson and R. Kalish, Phys. Rev. Lett. 69 (1992) 2991

[10] M.B.H. Breese, D.N. Jamieson, P.J.C. King, Material Analysis Using a Nuclear Microprobe, John Wiley and Sons Inc., New York, 1996.

[11] P. Olivero, S. Rubanov, P. Reichart, B. C. Gibson, S. T. Huntington, J. R. Rabeau, A. D. Greentree, J. Salzman, D. Moore, D. N. Jamieson, S. Prawer, Diamond Relat. Mater 15, (2006) 1614.

[12] P. Olivero, S. Rubanov, P. Reichart, B. C. Gibson, S. T. Huntington, J. Rabeau, A. D. Greentree, J. Salzman, D. Moore, D. N. Jamieson, S. Prawer, Adv. Mater. 17 (20), (2005) 2427-2430

[13] M. P. Hiscocks, K. Ganesan, B. C. Gibson, S. T. Huntington, F. Ladouceur, S. Prawer, Opt. Express 16 (24), (2008) 19512-19519

[14] S. Lagomarsino, P. Olivero, F. Bosia, M. Vannoni, S. Calusi, L. Giuntini, and M. Massi, Phys. Rev. Lett. 105, (2010) 233903

[15] I. Bayn, B. Meyler, A. Lahav, J. Salzman, R. Kalish, B. A. Fairchild, S. Prawer, M. Barth, O. Benson, T. Wolf, P. Siyushev, F. Jelezko, J. Wrachtrup, Diamond Relat. Mater. 20, (2011) 937-943

[16] J. C. Lee, I. Aharonovich, A. P. Magyar, F. Rol, E. L. Hu, Opt. Express 20 (8), (2012) 8891

[17] B. R. Patton, P. R. Dolan, F. Grazioso, M. B. Wincott, J. M. Smith, M. L. Markham, D. J. Twitchen, Y. Zhang, E. Gu, M. D. Dawson, B. A. Fairchild, A. D. Greentree, S. Prawer, Diamond Relat. Mater. 21, (2012) 16-23

[19] M. Liao, S. Hishita, E. Watanabe, S. Koizumi, Y. Koide, Adv. Mater. 22, (2010) 5393-5397

[20] M. K. Zalalutdinov, M. P. Ray, D. M. Photiadis, J. T. Robinson, J. W. Baldwin, J. E. Butler, T. I. Feygelson, B. B. Pate, B. H. Houston, Nano Lett. 11, (2011) 4304-4308

[21] S. Prawer, A. D. Devir, L. S. Balfour, R. Kalish Appl. Opt. 34, (1995) 636-640

[22] A. V. Karabutov, V. G. Ralchenko, I. I. Vlasov, R. A. Khmelnitsky, M. A. Negodaev, V. P. Varnin, I. G. Teremetskaya, Diamond Relat. Mater. 10, (2001) 2178-2183

[23] P. J. Sellin, A. Galbiati, Appl. Phys. Lett. 87, 093502 (2005)

[24] J. Forneris, V. Grilj, M. Jakšić, A. Lo Giudice, P. Olivero, F. Picollo, N. Skukan, C. Verona, G. Verona-Rinati, E. Vittone, Nucl. Instrum. Methods Phys. Res., sect. B 306, (2013) 181-185

[25] P. Olivero, J. Forneris, M. Jakšić, Ž. Pastuović, F. Picollo, N. Skukan, E. Vittone Nucl. Instrum. Methods Phys. Res., sect B 269, (2011) 2340-2344

[25] F. Picollo, S. Gosso, E. Vittone, A. Pasquarelli, E. Carbone, P. Olivero, V. Carabelli, Adv. Mater., 25 (2013) 4696 – 4700

[26] http://www.jcnabity.com/

[27] J.F. Ziegler, M.D. Ziegler, J.P. Biersack, Nucl. Instr. Meth., sect. B 268 (2010) 1818.

[28] W. Wu and S. Fahy, Phys. Rev. B 49 (1994) 3030

[29] J.D. Cutnell and K.W. Johnson 2004 Resistivity of Various Materials in Physics (New York: Wiley)

[30] S. Gosso, M. Turturici, C. Franchino, E. Colombo, A. Pasquarelli, E. Carbone, V. Carabelli, J. Physiol. 592 (2014) 3215

[31] Y. Lin, R. Trouillon, M.I. Svensson, J.D. Keighron, A.S. Cans, A.G. Ewing, Anal. Chem. 84 (2012) 2949–2954

[32] K. Kisler, K.B. Kim, X. Liu, K. Berberian, Q. Fang, C.J. Mathai, S. Gangopadhyay, K.D. Gillis, M. Lindau, J Biomater Nanobiotechnol. 3 (2012) 243-253

**Figures and captions**

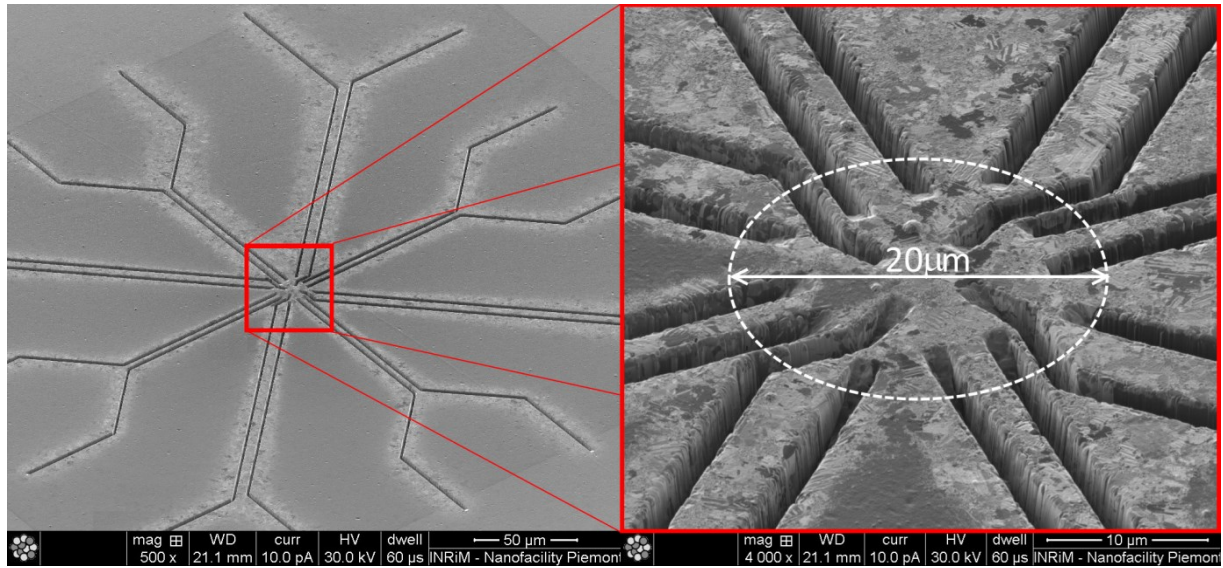

Fig. 1: SEM micrographs of the FIB micro-machined copper mask: a) overview of the whole milled region; b) zoom of the 300 μm$^2$ circular area where the 16 holes are converging; the slowly-thinning Cu ramps are visible.

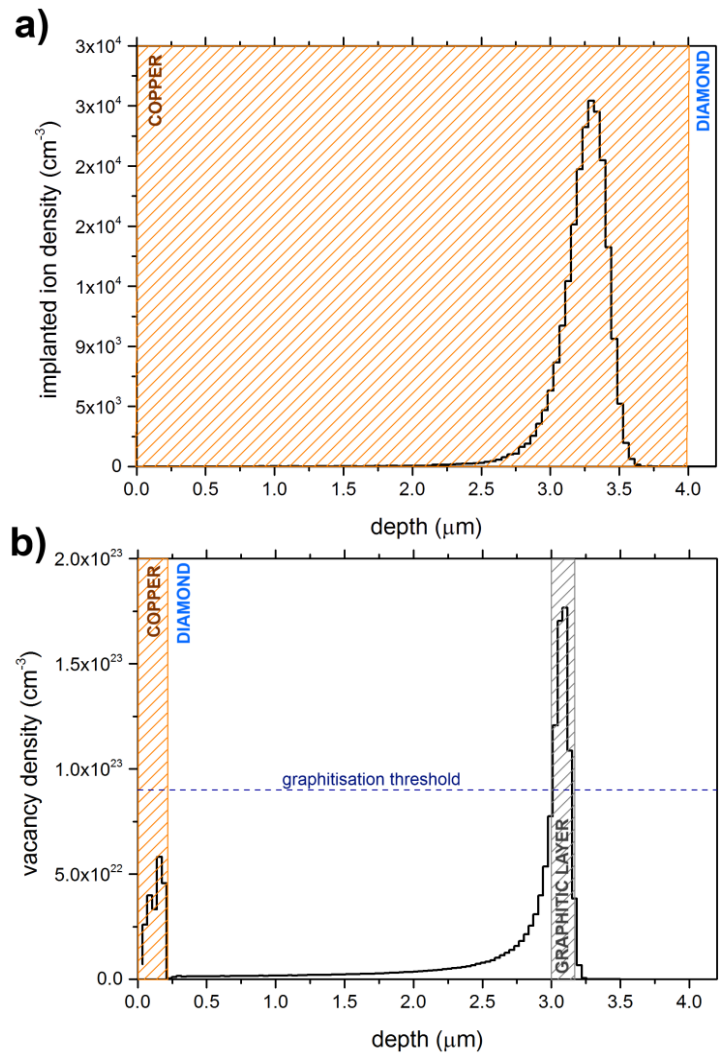

Fig. 2: SRIM Monte Carlo simulations of : a) ion distribution of 1.8 MeV He$^+$ implanted in a 4 μm thick copper layer: the Bragg pick is entirely located within the metal mask; b) vacancy density profile induced in diamond substrate covered with 200 nm Cu by 1.8 MeV He$^+$. The graphitization threshold is reported in dashed line. The graphitic region is highlighted in correspondence of the intersection of the Bragg peak with the graphitization threshold.

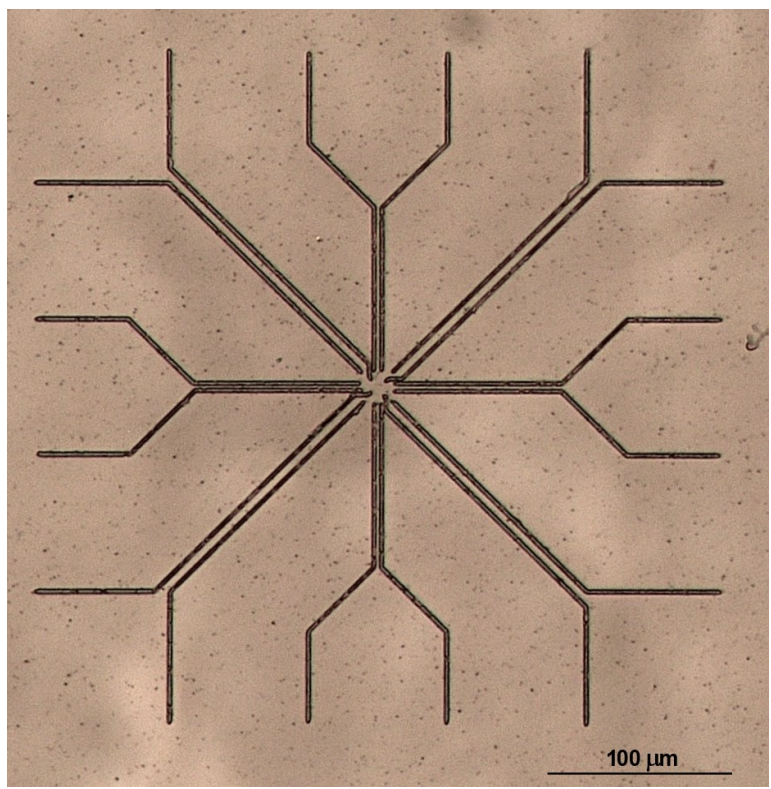

Fig. 3: Optical micrograh of the implanted diamond after thermal treatment. The graphic channels dimensions show a good correspondence with mask features.

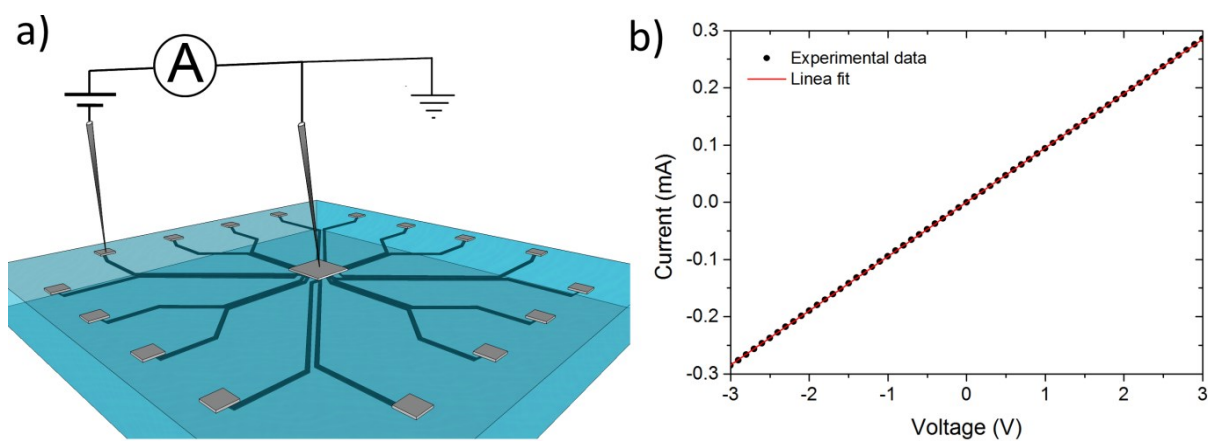

Fig. 4: a) Schematics of the experimental set-up realized for current-voltage measurements; b) example of an IV curve obtained by probing two pads at the endpoints of a single channel: the IV characteristic exhibits an ohmic behavior.

14